\documentclass[twocolumn,5p,10pt]{elsarticle}
\usepackage{amsmath, amsthm, amsfonts,amssymb}
\usepackage{lineno}
\usepackage[hidelinks]{hyperref}
\usepackage{graphicx}
\usepackage{subcaption}
\usepackage[format=hang]{caption}
\usepackage{natbib}
\biboptions{sort&compress}
\usepackage{color}
\usepackage{mathtools}

\newcommand{\B}{\mathfrak{B}}
\newcommand*{\END}{\hfill\ensuremath{\blacksquare}}
\newcommand{\proj}[2]{\pi_{#1}\left(#2\right)}

\theoremstyle{definition}
\newtheorem{definition}{Definition}
\newtheorem{theorem}{Theorem}
\newtheorem{problem}{Problem}
\newtheorem{example}{Example}
\newtheorem{corollary}[theorem]{Corollary}

\theoremstyle{remark}

\newtheorem{lemma}[theorem]{Lemma}
\newtheorem{remark}{Remark}

\makeatletter
\providecommand{\bigsqcap}{%
	\mathop{%
		\mathpalette\@updown\bigsqcup
	}%
}
\newcommand*{\@updown}[2]{%
	\rotatebox[origin=c]{180}{$\m@th#1#2$}%
}
\makeatother

\journal{Elsevier}

\begin{document}
	
\begin{frontmatter}

\title{Behavioural Approach to Distributed Control of Interconnected Systems}

\author[unsw]{Yitao Yan}
\ead{y.yan@unsw.edu.au}
\author[unsw]{Jie Bao}
\ead{j.bao@unsw.edu.au}
\author[ualb]{Biao Huang}\ead{biao.huang@ualberta.ca}

\address[unsw]{School of Chemical Engineering, UNSW Sydney, NSW 2052, Australia}
\address[ualb]{Department of Chemical and Materials Engineering, University of Alberta, AB T6G2G6, Edmonton, Canada}

\begin{abstract}
This paper formulates a framework for the analysis and distributed control of interconnected systems from the behavioural perspective. The discussions are carried out from the viewpoint of set theory and the results are completely representation-free. The core of a dynamical system can be represented as the set of all trajectories admissible through the system and interconnections are interpreted as constraints on the choice of trajectories. We develop a structure in which the interconnected behaviour can be directly built from the behaviours of the subsystems in an explicit way without any presumed forms of representations. We show that the interconnected behaviour can also be fully obtained from local observations of the subsystem. Furthermore, we develop the necessary and sufficient conditions for the existence of distributed controller behaviours and their explicit construction. Due to the entirely representation-free nature of this framework, it unites various representations and descriptions of features of dynamical systems (e.g. models, dissipativity, data, etc.) as behaviours, allowing for the formation of a unified platform for the analysis and distributed control for interconnected systems.
\end{abstract}


\end{frontmatter}


\section{Introduction}
\begin{figure}
	\centering
	\includegraphics[width=0.9\linewidth]{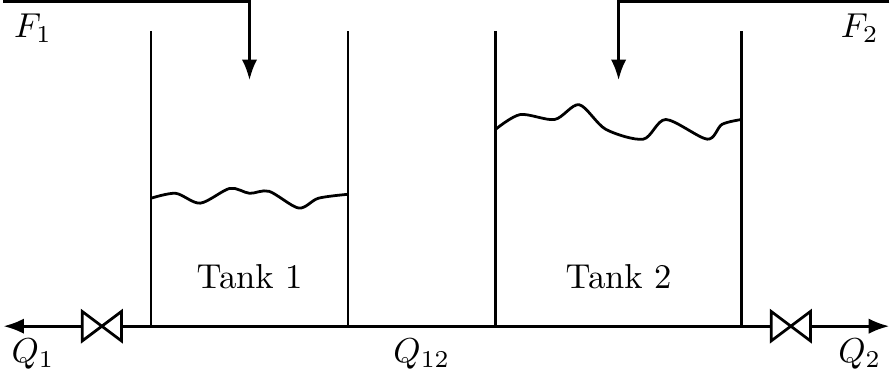}
	\caption{A Double-tank System}
	\label{fig:fig-twotank}
\end{figure}

The advancement of technology has made rapid collection and huge storage of data of large-scale, complex interconnected systems possible. The data sets contain rich information of the dynamics of these systems and as a result, a new paradigm of data-enhanced operations and data-driven control is emerging \cite{Stanley:2018}. However, the complex dynamics caused by the incredibly convoluted interconnections among subsystems pose grand challenges to the understanding of the system dynamics. To begin with, the dynamics of each subsystem can be vastly different from when it is a stand-alone system because it is always under the constraints posed by interconnections. While the changes in dynamics can be relatively easily captured in model-based approaches with models describing the dynamics, it is less so with data-driven methods. Furthermore, the characterisation of input and output is rather difficult for the interconnecting variables because the ``directions'' of their flows are unclear. To illustrate this point, consider a simple double-tank system depicted in Figure \ref{fig:fig-twotank}, in which the liquid in the two tanks are maintained at a relatively steady levels against possible disturbances in $F_1$ and $F_2$ through the manipulation of exit flowrates $Q_1$ and $Q_2$. The interconnecting flow $Q_{12}$, although theoretically measurable, cannot be manipulated in any way. While it is possible to construct an approximate model for this system from first principle, it is not clear whether $Q_{12}$ should be treated as the input or output for each tank and the direction of the flow  $Q_{12}$ depends on the liquid levels in both tanks. If the two tanks are two ``black boxes'' instead, the data set is the best way to describe the dynamics of the system because any empirical models can at best describe the dynamics of a system as good as the data set. However, data obtained for each ``box'' is always with the presence of interconnection, and the complexity of the dynamics escalates rapidly as the number of subsystems increases. Additionally, the concept of input/output becomes more vague because the complex interactions among subsystems may change the direction of information flow at any time.

\begin{figure}
	\centering
	\includegraphics[width=0.9\linewidth]{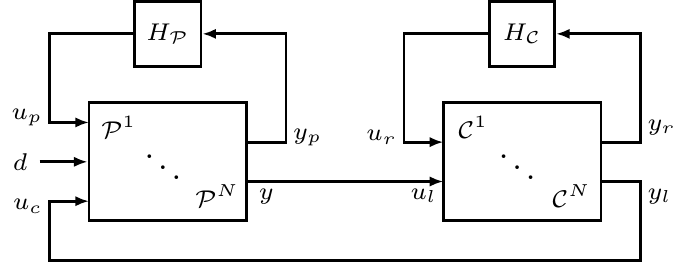}
	\caption{A Distributed Control Structure}
	\label{fig:fig-plantwideconventional}
\end{figure}

The complex dynamics due to interconnection also add considerable difficulty to the effective control of an interconnected system. With a good balance among complexity, performance and flexibility, distributed control strategy is often a preferred choice to control such a system \cite{Yan:2019}. A typical distributed control system is depicted in Figure \ref{fig:fig-plantwideconventional}, in which a network of dynamical systems are controlled by a network of controllers to achieve global control requirements in a flexible and robust manner. Currently, most efforts are focused on data analytics and modelling from a process database, and control design is based on these empirical models. However, the accuracy of these models can at most describe the systems as good as the original data sets. Furthermore, models obtained this way are inherently erroneous and the convoluted interconnections may well magnify such errors, leading to significant deterioration of control performance. The reason for such unavoidable sacrifice is because models are placed as the central role in defining a dynamical system while it is in fact not. This calls for a new way of thinking: a model is only to summarise some of the characteristics of a dynamical system rather than to define it. We need to analyse dynamical systems from a different angle and put the trajectories admissible through the system as the central role.

Initially proposed by Willems \cite{Willems:1991}, behavioural systems theory views a dynamical system as a set of functions mapped from a time axis to a signal space, or more commonly known as trajectories. This set, called the behaviour, is the centre of a dynamical system. As such, the theory views a dynamical system from a set-theoretic point of view. Analogous to the very nature of set theory that a set is defined by its elements, \emph{the trajectories that are admissible through a dynamical system define the system}. The behavioural approach provides not only a fresh viewpoint fundamentally different from model-based analysis, but also an excellent way in dealing with interconnections \cite{Willems:1997}. It does not distinguish between input and output but treats them as a single set of variables for the system, and interconnection of two dynamical systems is the sharing of trajectories between the two systems. As a result, the interconnected behaviour is simply the common trajectories between the two systems. This gives a flexible and scalable representation of an interconnected system, in that an additional subsystem integrated is essentially an additional set of constraints on the existing behaviour. With this view, control can be viewed as interconnection and controllers are essentially restricting the set of behaviour that can happen in the to-be-controlled systems.

A behavioural set admits different types of representations and each representation describes the set from a different perspective. Most notably, behaviours that are linear and time-invariant (LTI) are well studied and the relevant theory fits into the classical linear systems theory perfectly \cite{Polderman:1998}. Furthermore, the concept of dissipative dynamical systems is also well-developed both as a property of an existing dynamical system \cite{Willems:1972,Willems:1998,Tippett:2013a,Tippett:2014} and as a dynamical system itself with a ``dissipative behaviour'' \cite{Willems:2007a}. In terms of control design, controller synthesis for LTI behaviour with various representations have been discussed in \cite{Willems:2002,Trentelman:2002,Trentelman:2011} for stand-alone systems, and in \cite{Tippett:2013a,Tippett:2014,Yan:2019} for interconnected systems. In \cite{Willems:2005}, a condition to represent a finite-length LTI behaviour using persistently exciting trajectories has been given, and was modified to a more relaxed condition in \cite{Markovsky:2020}. Recent years have witnessed a proliferation of developments along this line in both analysis and control of LTI systems based on data \cite{Markovsky:2008,Maupong:2017,dePersis:2019,Romer:2019,Coulson:2019,Berberich:2020,vanWaarde:2020,Berberich:2020a,Koch:2020} as well as some extensions to a certain class of nonlinear systems \cite{Strasser:2020}. However, to the best of the authors' knowledge, a systematic, completely representation-free framework for the analysis and control is yet to be developed even for stand-alone systems, let alone interconnected ones. As discussed above, the observable behaviour of each subsystem is always under the influence by the complex interconnection among subsystems. The following questions then naturally arise: \emph{Are these restricted subsystem behaviours (i.e., behaviours of subsystems restricted by interconnections) sufficient for the analysis and control of interconnected systems? How to design such distributed controllers?}

In this paper, we develop a systematic approach to the analysis and distributed control of interconnected systems in an entirely representation-free fashion. By the novel definition of system network as a dynamical system with its own behaviour and through the use of the projection operation, we show that it is enough to determine the complete behaviour (i.e., all possible trajectories admissible through the system) of the interconnected system from the restricted behaviours of its subsystems or from the complete behaviours of some of the subsystems and the restricted behaviours of the rest. We then give necessary and sufficient conditions for the existence of the desired controlled behaviour that is both admissible through the system network and implementable through the controller network and we subsequently construct the behaviour of the distributed controllers. Following the rationale of behavioural systems theory, the control design procedure treats all variables equally without any assumption of causality, hence it is no longer an inverse problem. Furthermore, while the design requires rather delicate description of the controlled and controller behaviours, it is philosophically simple and intuitive. It should be pointed out that if all subsystems were described by models, then the proposed framework quickly reduces to the classical approaches such as $\mathcal{H}_{\infty}$ control \cite{Trentelman:1999a,Trentelman:2002} and dissipativity-based control \cite{Tippett:2014,Yan:2019}. At another extreme, for example, if all possible trajectories were given, then a \emph{na\"ive} realisation of the proposed framework is through brute force pattern matching. The proposed approach therefore builds a unified framework for distributed control design for interconnected systems whose subsystems admit a variety of, or even a mixture of, representations of their behaviours.

The rest of this paper is organised as follows: preliminary information about set operations and behavioural systems theory is introduced in Section \ref{sec:preliminaries}, the various ways of constructing the behaviour of an interconnected system through the projection operation is illustrated in Section \ref{sec:representation}, conditions to verify the existence of a desired controlled behaviour implementable through the controller network and the synthesis of the behaviours of distributed controllers  are presented in Section \ref{sec:controldesign}, and we conclude the paper in Section~\ref{sec:conclusion}. 

\textbf{Notations.} We denote the generic variable of a space $\mathbb{W}$ by $w$ and its dimension by $\mathrm{w}$. For an interconnected system, we denote the $j$th element of the variables in the $i$th subsystem as $w_j^i$ and its respective space as $\mathbb{W}_j^i$. The omission of subscript means that the focus is on all variables in the $i$th subsystem and the omission of superscript means that the interest is on the internal dynamics of a subsystem.  $\mathbb{Z}_N^+$ denotes the set of all positive integers less than or equal to $N$. We use the conventional $\cap$, $\cup$ and $\setminus$ to denote set intersection, union and difference, respectively. The Cartesian product of two sets $A$ and $B$, with their elements denoted by $a$ and $b$, respectively, is defined as $A\times B=\left\{(a,b)\mid a\in A \text{ and } b\in B\right\}$, which is neither commutative nor associative. For $N$ sets $A^1, \ A^2, \ \ldots, \ A^N$, we define
\begin{equation*}
	\bigtimes_{i=1}^{N}A^i\coloneqq A^1\times A^2\times\ldots\times A^N.
\end{equation*}

\section{Preliminaries}\label{sec:preliminaries}
\subsection{Behavioural Systems Theory}
In behavioural system theory, a dynamical system is viewed as a triple $\Sigma=(\mathbb{T},\mathbb{W},\B)$ where $\mathbb{T}$ is the time axis, $\mathbb{W}$ is the signal space and $\B\subset\mathbb{W}^\mathbb{T}$ is the behaviour, which contains the set of trajectories $w:\mathbb{T}\rightarrow\mathbb{W}$ admissible through the system \cite{Willems:1991}. The generic variable $w$ of this system is called the \emph{manifest} variable, which contains all variables of interest such as exogenous inputs and outputs. However, input and output are not \emph{a priori} distinguished from each other in $w$. Among the elements of a manifest variable, there is a set of variables called \emph{free variables}. For a dynamical system $\Sigma=\left(\mathbb{T},\mathbb{W}_1\times\mathbb{W}_2,\B\right)$, $w_1$ is said to be free if for all $w_1\in\mathbb{W}_1^\mathbb{T}$, there exists a $w_2\in\mathbb{W}_2^\mathbb{T}$ such that $(w_1,w_2)\in\B$, i.e. the set of possible trajectories for it is $\mathbb{W}_1^\mathbb{T}$ \cite{Polderman:1998}. Free variables include all exogenous inputs such reference and disturbance. If all variables in $w_1$ are free variables while none of the variables in $w_2$ are, then $\left(w_1,w_2\right)$ defines an input/output partition of $\B$. Other than the manifest variable, the system may also contain auxiliary variables $\ell$ called \emph{latent variables} that aid the description of a dynamical system (e.g., state variable in the classical state-space representation). In such a case, the \emph{full} system is the quadruple $\Sigma^{full}=(\mathbb{T},\mathbb{W},\mathbb{L},\B^{full})$ where $\B^{full}\subset(\mathbb{W}\times\mathbb{L})^\mathbb{T}$ and the \emph{manifest behaviour} $\B$ can be obtained as $\B=\left\{w\mid \exists\ell, (w,\ell)\in\B^{full}\right\}$. As an example, in the double-tank system depicted in Figure \ref{fig:fig-twotank}, variables $F_1$, $F_2$, $Q_1$, $Q_2$ and $Q_{12}$ are manifest variables, among which $F_1$ and $F_2$ are free variables, while the liquid levels in the two tanks can be seen as latent variables.

While a behaviour is in essence a set of trajectories, it can be represented in various ways and each description reveals the insights of a dynamical system from a different perspective. Here we give two examples of the well-known representations. 

\begin{enumerate}
	\item \emph{Model-based Difference Systems}, in which $\B$ is represented by a recursive difference model $f\left(\hat{w}(k),k\right)=0$, where $\hat{w}(k)=\mathrm{col}(\ldots,w(k-1),w(k),w(k+1),\ldots)$. $\B^{full}$ is represented by $f^{full}(\hat{w}(k),\hat{\ell}(k),k)=0$, where $\hat{\ell}(k)$ is defined similar to $\hat{w}(k)$.
	\item \emph{Data Banks}, in which $\B$ is entirely described by a set of data (e.g., trajectories of $w$). In this case, due to the finite number of trajectories with finite length, it only represents partial behaviour up to a certain length. If the stored trajectories have length $T$, then they can partially represent the finite-length behaviour $\B_{|[1,T]}=\left\{w\mid\exists w'\in\B, w(k)=w'(k), \forall k\in\mathbb{Z}_T^+\right\}$.
\end{enumerate}
It is seen here that by letting trajectories rather than the representations be the definitions of dynamical systems, we have the freedom of choosing our perspectives in viewing the dynamical systems and thus can capture much richer characteristics of them. A note to make here is that a dynamical system does \emph{not} necessarily admit a ``dynamic'' representation \cite{Polderman:1998}. For example, a proportional-only state feedback controller $u=-Kx$ obtained in a standard linear-quadratic control design has a static representation, but it only reveals the static relationship among variables. The behaviour of the system is still a set of trajectories that evolves with time, hence \emph{dynamic}.

Another merit of this framework is its insights in understanding interconnected systems, particularly for data-driven control due to the representation-free nature of the framework. Instead of viewing interconnections as signals flowing from one system to another, they are viewed as variables sharing the same trajectories. The behaviour of the interconnecting variable is hence the set of trajectories admissible through both systems. This is one of the key concepts in this framework: all trajectories are \emph{already contained} within the dynamical system and interconnection is restricting the possible choices of trajectories rather than forging new ones \cite{Willems:1997,Polderman:1998}. An interconnection is called a full interconnection if all variables between two variables are shared and a partial interconnection if a part of its variables are shared. All partial interconnections can be augmented into full interconnections by viewing variables from the other system that are not interconnected as free variables \cite{Willems:1997}. In this way, interconnection is simply excluding trajectories that are not admissible through any of the interconnecting systems. This view, adding to the representation-free nature of the behaviours themselves, allows for a systematic and generic approach to the analysis of complex interconnected dynamics.

\section{Representation of Interconnected Systems}\label{sec:representation}
In this section, we propose a new structure to obtain interconnected behaviour directly from the behaviours of the subsystems. We begin by introducing the concept of abstracting the network of an interconnected system as a dynamical system, thereby making it a stand-alone object with its own trajectories instead of being a feature of the interconnected system. This abstraction is quite useful in practice. For example, in the design of a large-scale chemical process, a flexible interconnection scheme is often implemented so that the plant is able to manufacture a variety of products to suit different needs. The dynamics of the process differ dramatically from each other with different interconnections but what is actually happening is that the network itself is dynamical. With this viewpoint, we give an explicit representation of the interconnected behaviour in a general form built up from its components without assuming any form of representation of the local behaviours.
\subsection{Network as a Dynamical System}
As proposed by \cite{Willems:1991}, interconnection of dynamical systems can be thought of as variable sharing, and henceforth one of the two variables interconnected together can be eliminated. This procedure yields a compact representation of an interconnected system and shows clearly what variables are left unconnected (i.e., the free variables). However, after this process, the membership of the interconnected variables to the subsystems becomes ambiguous: one variable is shared between two subsystems while it is in fact two distinct variables that happen to share the same value. It is only when \emph{all} variables of the interconnected subsystems are shared variables that the ambiguity disappears. We therefore propose a structure in which all subsystems are isolated but sharing all of their variables to a central system which can be seen as a generalised ``topology''. To explain the rationale, we lead in with an example. 
\begin{figure}
	\centering
	\begin{subfigure}{0.45\textwidth}
		\centering
		\includegraphics[width=\linewidth]{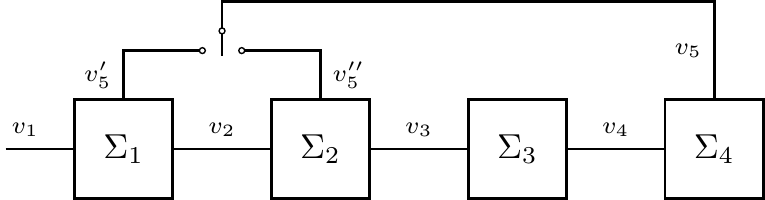}
		\caption{}
		\label{fig:intercon}
	\end{subfigure}
	\begin{subfigure}{0.45\textwidth}
		\centering
		\includegraphics[width=0.9\linewidth]{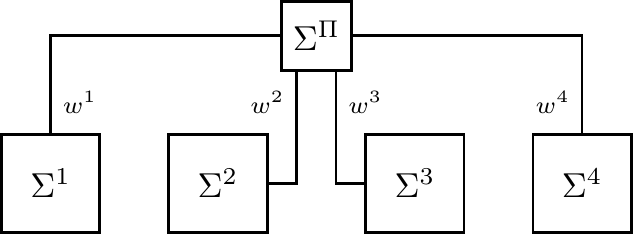}
		\caption{}
		\label{fig:intercon2}
	\end{subfigure}
	\caption{Four Systems with a Switching Network}
\end{figure}
\begin{example}
	Consider a network depicted in Figure \ref{fig:intercon}, in which four dynamical systems are interconnected in a network with a switch. Based on the outcome of the rest of the plant, $v_5$ can connect with either $v_5'$ or $v_5''$. Assuming $v_5$, $v_5'$ and $v_5''$ are of the same dimension, the interconnected system can be represented as $\Sigma=(\mathbb{T},\mathbb{V}_1\times\mathbb{V}_2\times\mathbb{V}_3\times\mathbb{V}_4\times\mathbb{V}_5\times\mathbb{V}_5,\B^{int})$ where
	\begin{equation*}
		\begin{split}
			\B^{int}&=\left\{(v_1,v_2,v_3,v_4,v_5,v_5'')\mid (v_1,v_2,v_5)\in\B^1,\right.\\
			&\qquad \left.(v_2,v_3,v_5'')\in\B^2,(v_3,v_4)\in\B^3,(v_4,v_5)\in\B^4\right\}
		\end{split}
	\end{equation*}
	if $v_5$ is interconnected with $v_5'$ and
	\begin{equation*}
		\begin{split}
			\B^{int}&=\left\{(v_1,v_2,v_3,v_4,v_5,v_5')\mid (v_1,v_2,v_5')\in\B^1,\right.\\
			&\qquad \left.(v_2,v_3,v_5)\in\B^2,(v_3,v_4)\in\B^3,(v_4,v_5)\in\B^4\right\}
		\end{split}
	\end{equation*}
	if $v_5$ is interconnected with $v_5''$. 
	
	While an insightful description of the interconnected behaviour, representing the behaviour in this slightly over-compacted fashion creates two obstacles towards the analysis of the system. Firstly, due to variable elimination, only one variable is used to represent variables that originally come from several different systems, making it difficult to construct the interconnected behaviour from that of its subsystems directly and explicitly. A more natural representation is that each subsystem has its own manifest variables and several of them ``happen to'' coincide during interconnection. Secondly, for different networks, variables shared among subsystems are different. If the above method of representation were adopted, a new representation would be needed every time the interconnection changes. What is actually changing is the network itself, and pushing the ``dynamics'' of the network into the subsystems makes the analysis of the interconnected behaviour more complicated. It is therefore reasonable to treat the network as a dynamical system itself with its own behaviour.
	\begin{figure}
		\centering
		\includegraphics[width=0.9\linewidth]{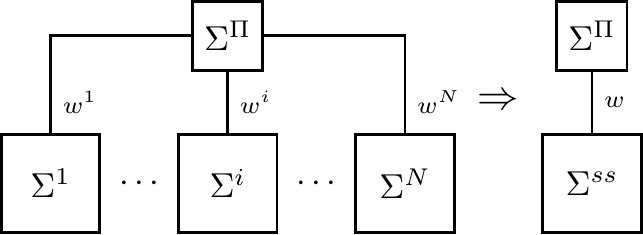}
		\caption{Network Viewed as a Dynamical System}
		\label{fig:interconnew}
	\end{figure}
	With this thinking, the interconnected system in Figure \ref{fig:intercon} can be equivalently depicted as in Figure \ref{fig:intercon2}, in which four (isolated) dynamical systems $\Sigma^i=(\mathbb{T},\mathbb{W}^i,\B^i)$ are ``plugged'' into a dynamical system $\Sigma^\Pi=(\mathbb{T},\mathbb{W}^1\times\mathbb{W}^2\times\mathbb{W}^3\times\mathbb{W}^4,\B^\Pi)$ that is the network. By setting $w=\mathrm{col}(w^1,w^2,w^3,w^4)$, the interconnected system can be described as $\Sigma=(\mathbb{T},\mathbb{W}^1\times\mathbb{W}^2\times\mathbb{W}^3\times\mathbb{W}^4,\B)$ where $$\B=\left\{w\mid w^i\in\B^i, i\in\mathbb{Z}_4^+ \text{ and } w\in\B^\Pi\right\}.$$ In this way, the interconnected system can be constructed directly from its components. \END
\end{example}

Note that this description of a interconnected behaviour can be generalised into arbitrary numbers of subsystems. This leads to the definition of a network as a dynamical system.
\begin{definition}\label{definition:network}
	The \emph{network} of an interconnected system consisting of $N$ subsystems, with the $i$th subsystem denoted as $\Sigma^i=(\mathbb{T},\mathbb{W}^i,\B^i)$, is itself a dynamical system
	\begin{equation}
		\Sigma^\Pi=\left(\mathbb{T},\bigtimes_{i=1}^{N}\mathbb{W}^i,\B^\Pi\right)
	\end{equation}
	where $\B^\Pi$ is the \emph{network behaviour}.
\end{definition}

The idea of giving the network a representation has already appeared in literature. In \cite{Willems:1972}, the network was represented by a static interconnection function, and in \cite{Rojas:2009,Tippett:2013,Yan:2019}, it was represented by a static LTI system using input/output representation. Definition \ref{definition:network} encompasses these cases and generalises the network to be a dynamical system on its own, which allows a more systematic description of an interconnected system with an arbitrary and probably time varying topology. It is, however, important to note the difference between the network behaviour defined in Definition \ref{definition:network} and the topology matrix/interconnecting function in the literature: the network behaviour has it \emph{own} set of behaviour whereas those in the literature are entirely defined by the interconnection inputs and outputs. To name but one distinct difference, if all subsystems were isolated, the topology matrix defined in \cite{Yan:2019}, or $H_\mathcal{P}$ in Figure \ref{fig:fig-plantwideconventional}, for example, would be 0, whereas the behaviour defined in Definition \ref{definition:network} would be $\B^\Pi=\mathbb{W}^\mathbb{T}$ where $\mathbb{W}=\bigtimes_{i=1}^{N}\mathbb{W}^i$, which means that all mappings from the time axis to the signal space are admissible in $\Sigma^\Pi$. A closer observation reveals that in this particular case, the former is in fact a representation of $\B^\Pi$. In this way, variables with no physical interconnection can also be interpreted as ``interconnected'' with the network. In the next few sections, we will show that not only does the proposed structure provide a clean and flexible representation of the interconnected system that can be constructed from its subsystems explicitly, but the network behaviour is also the key to the construction of interconnected behaviour in various ways.
\subsection{The Interconnected Behaviour}
As illustrated above, the proposed structure allows for the representation of all interconnections as full interconnections. A result of this is that the behaviour of a dynamical system formed by the interconnection of two subsystems, denoted by $\Sigma=\Sigma^1\wedge\Sigma^2$, can be directly constructed as $\B=\B^1\cap\B^2$. If $\Sigma^1$ and $\Sigma^2$ have two distinct signal spaces $\mathbb{W}^1$ and $\mathbb{W}^2$, the behaviour of $\Sigma$ can be more clearly represented by $\B=\B^1\times\B^2$. We denote the interconnected system constructed through Cartesian product as $\Sigma=\Sigma^1\sqcap\Sigma^2$. Note that the relationship between $\sqcap$ and $\wedge$ is similar to that between $\times$ and $\cap$. It is also straightforward to construct the interconnected behaviour from the relationship among different components if all behaviours were fully known by replacing $\Sigma$ with $\B$, $\wedge$ with $\cap$ and $\sqcap$ with $\times$. For example, a dynamical system $\Sigma=\left(\Sigma^1\sqcap\Sigma^2\right)\wedge\Sigma^3$ has behaviour $\B=\left(\B^1\times\B^2\right)\cap\B^3$.

As shown in Figure \ref{fig:interconnew}, assuming that the plant consists of $N$ subsystems with the $i$th subsystem denoted as $\Sigma^i=(\mathbb{T},\mathbb{W}^i,\B^i)$, then all subsystems can be written together compactly as a large system $\Sigma^{ss}$ with a set of isolated subsystems, i.e.,
\begin{equation*}
	\Sigma^{ss}=\bigsqcap_{i=1}^N\Sigma^i=\left(\mathbb{T},\bigtimes_{i=1}^{N}\mathbb{W}^i,\bigtimes_{i=1}^{N}\B^i\right).
\end{equation*} 
The final interconnection is the interconnection of $\Sigma^{sys}$ with $\Sigma^\Pi$. Since they have exactly the same signal space and they have full interconnection, the final interconnected system can be obtained as $\Sigma=\Sigma^{sys}\wedge\Sigma^\Pi$, i.e.,
\begin{equation}\label{eq:interconsystem}
	\Sigma=\left(\bigsqcap_{i=1}^N\Sigma^i\right)\wedge\Sigma^\Pi=\left(\mathbb{T},\bigtimes_{i=1}^{N}\mathbb{W}^i,\B\right),
\end{equation}
where
\begin{equation}\label{eq:interconbehaviour}
	\B=\left(\bigtimes_{i=1}^{N}\B^i\right)\cap\B^\Pi.
\end{equation}

Note that in the last part we still adopt the variable elimination procedure by naming the manifest variable of the network $\Sigma^\Pi$ as $w$ (the collection of all manifest variables in the subsystems) because $\Sigma^{sys}$ and $\Sigma^\Pi$ are indeed sharing all variables. In this way, the manifest variables of each subsystem is defined unambiguously because the ``sharing of variables'' happen inside the network $\Sigma^\Pi$. 

While \eqref{eq:interconbehaviour} gives a \emph{representation} of the interconnected behaviour $\B$, it is \emph{not} how $\B$ is defined. In other words, \eqref{eq:interconbehaviour} should be interpreted as an equivalence relationship rather than a definition of $\B$. This means that $\B$ admits other representations as well. In fact, in many cases, the complete behaviours of the subsystems $\B^i$ are hardly available. For example, if only data sets were available, the data set of local measurements of a subsystem, say the $i$th one, is no longer $\B^i$, but rather the projection of the interconnected behaviour $\B$ onto the space of the behaviour of $w^i$ in a network configuration. In the next section, a detailed treatment of the projection operation will be carried out. 

\subsection{Construction of Interconnected Behaviour through Projection}\label{sec:projection}
In this section, we first provide relevant properties of the projection operation and then we show that it can be used to construct the exact interconnected behaviour from local observations. Since the interconnected behaviour may not be obtained directly through the complete behaviour of the subsystems, using \eqref{eq:interconbehaviour} to express the behaviour of \eqref{eq:interconsystem} may no longer be feasible. In such a case, the interconnected systems are only given in terms of their subsystems $\Sigma^i$ and the network $\Sigma^\Pi$ to show how they physically interconnect.

Projection is one of the key operations in relational algebra, the algebra of data sets \cite{Codd:1970}. We use this operation in this paper in the context of behaviour. Given a dynamical system $\Sigma=\left(\mathbb{T},\mathbb{W},\B\right)$, the projection of the behaviour $\B$ onto the space $\mathbb{W}_i^\mathbb{T}$ is a map $\pi_{w_i}:\mathbb{W}^\mathbb{T}\rightarrow\mathbb{W}_i^\mathbb{T}$ defined by
\begin{equation}\label{eq:projection}
	\begin{split}
		&\proj{w_i}{\B}=\left\{w_i\mid\exists \ell_j, j\in\mathbb{Z}_\mathrm{w}^+\setminus\{i\}, \right.\\ &\qquad\qquad\quad \ \left.(\ell_1,\ldots,\ell_{i-1},w_i,\ell_{i+1},\ldots,\ell_\mathrm{w})\in\B\right\}.
	\end{split}
\end{equation}
This map allows for the extraction of the set of trajectories of any specific manifest variable from $\B$. Obviously, if the dynamical system is one with latent variables $\Sigma^{full}=(\mathbb{T},\mathbb{W},\mathbb{L},\B^{full})$, then the manifest behaviour is given by $\B=\pi_w\left(\B^{full}\right)$. 

The projected behavioural set can be understood in two ways: from the point of view of the dynamical system $\Sigma$, $\proj{w_i}{\B}$ can be interpreted as the observation of all possible trajectories of $w_i$; from the point of view of the manifest variable $w_i$ itself, $\proj{w_i}{\B}$ can also be interpreted as a virtual ``dynamical system'' with manifest variable $w_i$ having full interconnection with another virtual ``dynamical system'' with the same manifest variable $w_i$.  In this view, all other manifest variables in $\Sigma$ are treated as latent variables(see Lemma \ref{lemma: projection}(i) for a mathematical representation). Since the main focus of this paper is on the latter interpretation, the definition of $\proj{w_i}{\B}$ in \eqref{eq:projection} deliberately uses $\ell_j$ instead of $w_j$ to emphasise that the choices of $\ell_j$ may not be unique and that $w_j$ may be only one of the choices. If, however, the choices of $w_j$ are actually unique, then the system is said to be observable. In fact, we have the following definition.
\begin{definition}[Observability \cite{Willems:1991}]
	Given a dynamical system $\Sigma=(\mathbb{T},\mathbb{W},\B)$ with manifest variable $w$ partitioned as $w=(w_1,w_2)$, $w_1$ is said to be observable from $w_2$ in $\Sigma$ if $\{(w_1,w_2)\in\B, \  (w_1',w_2)\in\B\}\Rightarrow \{w_1=w_1'\}$ for all $w_2$.
\end{definition}
Using the projection operation, this definition shows that if $w_1$ is observable from $w_2$ in $\Sigma$, then for a given trajectory of $w_2\in\proj{w_2}{\B}$, there exists \emph{only one} trajectory of $w_1\in\proj{w_1}{\B}$ such that $(w_1,w_2)$ is an admissible trajectory in $\Sigma$. In other words, the complete behaviour of $w_2$ can be fully determined from that of $w_1$, although $w_1$ being observable from $w_2$ does not necessarily mean that each trajectory of $w_2$ corresponds to a distinct trajectory of $w_1$. It is perfectly possible for different trajectories of $w_2$ to have the same corresponding trajectory of $w_1$.

We now present the main result of this section. It claims that the behaviour of an interconnected system can be reconstructed from the projections of the behaviour onto the behaviour space of each subsystem as well as the network behaviour. 
\begin{theorem}\label{theorem:behaviourPW}
	Given an interconnected system \eqref{eq:interconsystem}, then
	\begin{enumerate}[(i)]
		\item assuming that behaviours $\B^i$ are not known but the projections on their manifest variables are known, the interconnected behaviour can be fully obtained as
		\begin{equation}\label{eq:interconbehaviourproj}
			\B=\left(\bigtimes_{i=1}^{N}\proj{w^i}{\B}\right)\cap\B^\Pi;
		\end{equation}
		\item assuming, without loss of generality, that the first $n$ behaviours are fully known while the rest only have information of the projections, the interconnected behaviour can be fully obtained as
		\begin{equation}\label{eq:interconbehaviourhybrid}
			\B=\left[\left(\bigtimes_{i=1}^{n}\B^i\right)\times\left(\bigtimes_{i=n+1}^{N}\proj{w^i}{\B}\right)\right]\cap\B^\Pi.
		\end{equation}
	\end{enumerate}
\end{theorem}
\begin{proof}
	See Appendix \ref{appx:proofbehaviourPW}.
\end{proof}

The first claim of Theorem \ref{theorem:behaviourPW} is that all of the projections of the interconnected behaviour, together with the network behaviour, determine the interconnected behaviour completely. This provides an insight to an interconnected system: each subsystem within an interconnected system contains a set of trajectories that can \emph{never} happen. Therefore, it is in fact not necessary to obtain the complete information of each subsystem. The behaviour of each subsystem \emph{as an integrated part of the interconnected system} is enough to determine the complete interconnected behaviour. The most interesting part of this statement is that we still need the network behaviour to construct the interconnected behaviour even though the projections already contain the network information. This can be understood from the property of the projection operation: by projecting $\B$ onto $(\mathbb{W}^i)^\mathbb{T}$, all manifest variables of other subsystems are viewed as latent variables with respect to $w^i$. As such, there may be trajectories that are not admissible through the interconnected system but indistinguishable from $w^i$. The network behaviour precisely eliminates this problem because any trajectory that is admissible in the interconnected system must be admissible through the network behaviour. In many cases, $\proj{w^i}{\B}$ can be obtained to a high level of completeness (e.g., with a large data bank of measured trajectories) and $\B^\Pi$ is essentially known completely, making it possible for data-driven control design under this framework. 

The second claim of Theorem \ref{theorem:behaviourPW}, on the other hand, is a more powerful result. It states that if an interconnected system contains several subsystems with fully known behaviours (e.g., behaviour described by models), then the complete interconnected behaviour can be fully obtained using these complete behaviours, the observations of the behaviours of the rest of the subsystems and the network behaviour. In other words, the proposed construction allows for a unified platform for a hybrid interconnected system whose subsystems can be described by deterministic representations, data sets, or both.

If the interconnected system contains only two subsystems, then Theorem \ref{theorem:behaviourPW} reduces to the corollary below.
\begin{corollary}\label{corollary:twosystemproj}
	Given a dynamical system 
	\begin{equation}\label{eq:twosys}
		\Sigma=\left(\Sigma^1\sqcap\Sigma^2\right)\wedge\Sigma^\Pi=(\mathbb{T},\mathbb{W}^1\times\mathbb{W}^2,\B),
	\end{equation}
	we have
	\begin{equation*}
		\begin{split}
			\B&=\left[\proj{w^1}{\B}\times\B^2\right]\cap\B^\Pi=\left[\B^1\times\proj{w^2}{\B}\right]\cap\B^\Pi\\
			&=\left[\proj{w^1}{\B}\times\proj{w^2}{\B}\right]\cap\B^\Pi.
		\end{split}
	\end{equation*}
\end{corollary}
\begin{proof}
	This is a direct result from Theorem \ref{theorem:behaviourPW} by setting $N=2$ and $n=1$.
\end{proof}

This result is a key stepping stone in the synthesis of controlled behaviour in the next section as the interconnected system and the distributed controllers can be seen as two large subsystems.

\section{Distributed Control Design}\label{sec:controldesign}
This section presents the procedure of obtaining the behavioural sets for distributed controllers that, when integrated into the system, yield the desired behaviour for the variables of interest. We provide necessary and sufficient conditions for the existence of the controlled behaviour, from which the controller behaviours can also be obtained.

\subsection{Control Structure and Problem Formulation}
Consider an interconnected system $\Sigma_p$ consisting of $N$ subsystems, where the $i$th subsystem is denoted as $\Sigma_p^i=\left(\mathbb{T},\mathbb{W}_p^i,\B_p^i\right)$. The subsystems are interconnected with a network $\Sigma_p^\Pi=\left(\mathbb{T},\mathbb{W}_p,\B_p^\Pi\right)$. As a result, the interconnected, uncontrolled behaviour can be constructed according to \eqref{eq:interconbehaviour} as
\begin{equation}\label{eq:plantbehaviour}
	\Sigma_p=\left(\bigsqcap_{i=1}^{N}\Sigma_p^i\right)\wedge\Sigma_p^\Pi=\left(\mathbb{T},\mathbb{W}_p,\B_p\right),
\end{equation}
where $\B_p$ can be constructed according to \eqref{eq:interconbehaviour}, \eqref{eq:interconbehaviourproj} or \eqref{eq:interconbehaviourhybrid}. Suppose that a set of $N_c$ controllers $\Sigma_c^j=\left(\mathbb{T},\mathbb{W}_c^j,\B_c^j\right)$ are employed to control $\Sigma_p$ and the controllers have their own network $\Sigma_c^\Pi=\left(\mathbb{T},\mathbb{W}_c,\B_c^\Pi\right)$. Then the interconnected controllers can be represented as an interconnected system $\Sigma_c=\left(\mathbb{T},\mathbb{W}_c,\B_c\right)$, where
\begin{equation}\label{eq:controllerbehaviour}
	\B_c=\left(\bigtimes_{j=1}^{N_c}\B^j_c\right)\cap\B_c^\Pi.
\end{equation}
Note that the number of controllers is not necessarily the same as that of the subsystems, nor does $w_c$ have any relationship with $w_p$ at this stage. 

\begin{figure}
	\centering
	\includegraphics[width=0.7\linewidth]{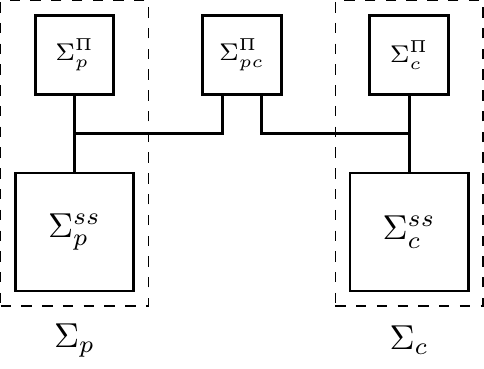}
	\caption{The Interconnected System Layout}
	\label{fig:Plantwide}
	\vskip -3mm
\end{figure}

When interconnecting the system with the distributed controllers, another network is needed. This network is defined as $\Sigma_{pc}^\Pi=\left(\mathbb{T},\mathbb{W}_p\times\mathbb{W}_c,\B_{pc}^\Pi\right)$. With these building blocks, a controlled system can be constructed as in Figure \ref{fig:Plantwide}. As is depicted, the controlled system can be viewed as the interconnection of two interconnected systems, which defines a latent variable dynamical system $\Sigma_{pc}^{full}=\left(\Sigma_p\sqcap\Sigma_c\right)\wedge\Sigma_{pc}^\Pi=\left(\mathbb{T},\mathbb{W}_p,\mathbb{W}_c,\B_{pc}^{full}\right)$ with full behaviour
\begin{equation}\label{eq:fullplanwide}
	\B_{pc}^{full}=(\B_p\times\B_c)\cap\B_{pc}^\Pi
\end{equation}
The controlled system can then be expressed as the triple $\Sigma_{pc}=\left(\mathbb{T},\mathbb{W}_p,\B_{pc}\right)$ where
\begin{equation}\label{eq:manifestplantwide}
	\B_{pc}=\proj{w_p}{\B_{pc}^{full}}
\end{equation}
and we say that $\B_{pc}$ is \emph{implemented} by the controllers through $w_c$ \cite{Willems:2002}. Note that by defining
\begin{equation*}
	\B_p'\coloneqq\left(\B_p\times\B_c^\Pi\right)\cap\B_{pc}^\Pi,
\end{equation*}
we have
\begin{equation}\label{eq:controldecentral}
	\B_{pc}^{full}=\B_p'\cap\left[\mathbb{W}_p^\mathbb{T}\times\left(\bigtimes_{j=1}^{N_c}\B^j_c\right)\right],
\end{equation}
which shows that the distributed control design is equivalent to decentralised control design for an augmented system with controller network and system-controller network integrated. By treating the networks as dynamical systems, they have their own behavioural sets and can thus be treated and rearranged like physical subsystems. 
\begin{remark}
	The proposed structure is general and encompasses a range of system configurations. As an example, for the system depicted in Figure \ref{fig:fig-plantwideconventional} with fixed topology, we have $w_p=\begin{bmatrix}
		y^T & y_p^T & u_p^T& u_c^T & d^T
	\end{bmatrix}^T$ and $w_c=\begin{bmatrix}
		y_l^T & y_r^T & u_l^T& u_r^T
	\end{bmatrix}^T$. The behaviours of $\Sigma_p^\Pi$ and $\Sigma_c^\Pi$ can then be described as kernel representations $\B_p^\Pi=\ker(\Pi_p)$ and $\B_p^\Pi=\ker(\Pi_c)$, respectively, where $\Pi_p$ and $\Pi_c$ are two matrices of proper dimensions (see \cite{Polderman:1998} for details about kernel representation). $\Sigma_{pc}^\Pi$, on the other hand, has manifest variable $w_{pc}=\begin{bmatrix}
		w_p^T & w_c^T
	\end{bmatrix}^T$ and behaviour $\B_{pc}^\Pi=\ker(\Pi_{pc})$, where $\Pi_{pc}$ is defined in similar way as $\Pi_p$. Note that these matrices are not describing the selection of process variables (as what $H_\mathcal{P}$ in Figure \ref{fig:fig-plantwideconventional} does), but rather a dynamical system with its own internal trajectories (i.e., behaviour), which can be described by the aforementioned representations. It will be shown in the coming section that these trajectories are what enables the control design.
\end{remark}

\begin{figure}
	\centering
	\includegraphics[width=\linewidth]{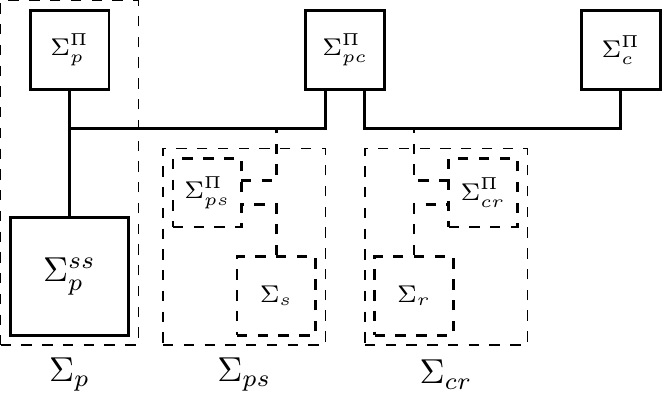}
	\caption{The Desired System Behaviour}
	\label{fig:Desired}
\end{figure}

The objective of control is to find the set of behaviour from the uncontrolled system such that all trajectories in the set meet certain specifications. These specifications can be formulated as a behavioural set $\B_{ps}$ imposed on all manifest variables $w_p$. Therefore, the control design aims to implement a subset of $\B_p\cap\B_{ps}$ through $w_c$. On the other hand, the controllers themselves may have restrictions and objectives such as control saturation and minimum gain requirement, which can also be formulated as a set of behaviour $\B_{cr}$ on the control variables $w_c$. Although for illustration purpose we assumed that the system and specifications share the same signal space, it is easy to formulate such a $\B_{ps}$ even if the requirements are specified otherwise. As depicted in Figure \ref{fig:Desired}, suppose the desired requirements are described by a set $\B_s\subset\mathbb{W}_s^\mathbb{T}$ with manifest variable $w_s$, then for $\B_s$ to be able to restrict $\B_p$, there must exist a network behaviour $\B_{ps}^\Pi\subset\left(\mathbb{W}_p\times\mathbb{W}_s\right)^\mathbb{T}$ such that for all $w_p\in\B_p$, there exists $w_s\in\B_s$ such that $(w_p,w_s)\in\B_{ps}^\Pi$. In this case, the set describing the requirements can be constructed as
\begin{equation*}
	\B_{ps}=\proj{w_p}{\left(\mathbb{W}_p^\mathbb{T}\times\B_s\right)\cap\B_{ps}^\Pi}.
\end{equation*}
Similarly, if the restrictions are imposed on the variable $w_r\in\B_r\subset\mathbb{W}_r^\mathbb{T}$, then there must exist a network behaviour $\B_{cr}$ linking $w_c$ to $w_r$. In such a case, $\B_{cr}$ can be constructed as
\begin{equation*}
	\B_{cr}=\proj{w_c}{\left(\mathbb{W}_c^\mathbb{T}\times\B_r\right)\cap\B_{cr}^\Pi}.
\end{equation*}
For the clarity of presentation, we will use the notations $\B_{ps}$ and $\B_{cr}$ and we assume that $\B_{ps}$ and $\B_{cr}$ always have the same signal spaces as $\B_p$ and $\B_c^\Pi$, respectively.

With these components, the control problem to be solved can be formulated as follows:

\begin{problem}\label{problem:problem}
	Given an interconnected dynamical system $\Sigma_p=\left(\mathbb{T},\mathbb{W}_p,\B_p\right)$ constructed according to \eqref{eq:plantbehaviour}, the control objective described by $\B_{ps}$, the controller network $\Sigma_c^\Pi=\left(\mathbb{T},\mathbb{W}_c,\B_c^\Pi\right)$ and the system-controller network $\Sigma_{pc}^\Pi=\left(\mathbb{T},\mathbb{W}_p\times\mathbb{W}_c,\B_{pc}^\Pi\right)$, design, if possible, a distributed control system with $N_c$ controllers $\Sigma_c^j=(\mathbb{T},\mathbb{W}_c^j,\B_c^j)$ such that
	\begin{enumerate}[1.]
		\item the controlled behaviour \eqref{eq:manifestplantwide} after interconnection of $\Sigma_p$ and $\Sigma_c$ (as shown in Figure \ref{fig:Plantwide}) satisfies
		\begin{subequations}\label{eq:objectives}
			\begin{align}
				\B_{pc}&\subset\B_p\cap\B_{ps} \label{eq:implementability}\\
				\proj{w_f}{\B_{pc}}&=\mathbb{W}_f^\mathbb{T} \label{eq:DOF},
			\end{align}
		\end{subequations}
		where $w_f$ denotes the free variable after interconnection;
		\item the resulting distributed controller $\Sigma_c=\left(\mathbb{T},\mathbb{W}_c,\B_c\right)$ with $\B_c$ described in \eqref{eq:controllerbehaviour} satisfies $\B_c\subset\B_{cr}$.
	\end{enumerate}
\end{problem}
The first objective is about the controlled behaviour. As specified in \eqref{eq:implementability}, the control design should result in the manifest controlled behaviour to be the subset of the uncontrolled system behaviour whose trajectories satisfy the requirements. Furthermore, as required by \eqref{eq:DOF}, the free variable $w_f$ (which normally contains exogenous inputs such as disturbances) should still be able to choose any trajectories it prefers after the integration of the controller network. The second objective relates to the physical constraints of the controllers: the resulting distributed controller must meet the restrictions specified by $\B_{cr}$, which can include constraints such as maximum range of the controller variables and economic cost involving them.

\subsection{Controller Behaviour Synthesis}
This section gives the main result of this paper: the construction of behaviours of the distributed controllers. We firstly explain, intuitively, the rationale of the control design. As stated in Problem \ref{problem:problem}, the given components are the subsystems $\Sigma_p^i$, the system network $\Sigma_p^\Pi$, the controller network $\Sigma_c^\Pi$ and the system-controller network $\Sigma_{pc}^\Pi$. The specifications on the system and the restriction on the controllers can be constructed as two virtual ``systems'' $\Sigma_{ps}=\left(\mathbb{T},\mathbb{W}_p,\B_{ps}\right)$ and $\Sigma_{cr}=\left(\mathbb{T},\mathbb{W}_c,\B_{cr}\right)$, respectively. By doing so, the two virtual systems can be integrated into the given components, resulting in a desired objective dynamical system as shown in Figure \ref{fig:Desired}. Obviously, the full behaviour of this system, denoted by $\B_d$, is
\begin{equation}\label{eq:desirelargest}
	\B_d\coloneqq\left[\left(\B_p\cap\B_{ps}\right)\times\left(\B_c^\Pi\cap\B_{cr}\right)\right]\cap\B_{pc}^\Pi.
\end{equation}
The projections of the behaviour of this dynamical system onto the spaces $\mathbb{W}_p^\mathbb{T}$ and $\mathbb{W}_c^\mathbb{T}$ give the largest possible set of behaviours for $w_p$ and $w_c$, respectively, that are admissible through the network. It is easy to see that $\proj{w_p}{\B_d}\subset\B_p\cap\B_{ps}$ according to Lemma \ref{lemma: projection}(i). Then, as shown in \eqref{eq:controldecentral}, since the controllers can be viewed as decentralised, the projection on the control variables of each controller $w_c^j$ gives the behavioural set for the corresponding controller $\Sigma_c^j$. Therefore, if the controllers with the aforementioned behavioural sets are integrated into the system depicted in Figure \ref{fig:Plantwide}, the resulting behaviour should, in some sense, resemble that shown in Figure \ref{fig:Desired}. 

We now rigorously formulate the above illustration with the following theorem. It shows that under certain conditions, the control design can indeed be carried out through this rationale, but both the largest possible set for the controlled behaviour and the resulting controller behaviours require much more delicate descriptions.

\begin{theorem}\label{theorem:controlsynthesis}
	The desired controlled behaviour \eqref{eq:objectives} exists and is implementable through a distributed control system described in Problem \ref{problem:problem} if and only if
	\begin{subequations}\label{eq:existence}
		\begin{align}
			\begin{split}\label{eq:exist1}
				&\proj{w_f}{\B_p\setminus\proj{w_p}{\B_{in}}\setminus\proj{w_p}{\B_{xi}}} \\
				&\hspace{9em}\subset\proj{w_f}{\B_{in}}\cup\proj{w_f}{\B_{xi}},
			\end{split}\\
			&\proj{w_f}{\B_p}=\mathbb{W}_f^\mathbb{T},\label{eq:exist2}
		\end{align}
	\end{subequations}
	where
	\begin{equation*}
		\begin{split}
			\B_{in}&=\left[\left(\proj{w_p}{\B_d}\setminus\proj{w_p}{\B_{ex}}\right)\times\left(\B_c^\Pi\cap\B_{cr}\right)\right]\cap\B_{pc}^\Pi,\\
			\B_{ex}&=\left[\proj{w_p}{\B_d}\times\proj{w_c}{\B_{out}}\right]\cap\B_{pc}^\Pi,\\
			\B_{out}&=\left[\left(\B_p\setminus\proj{w_p}{\B_d}\right)\times\left(\B_c^\Pi\cap\B_{cr}\right)\right]\cap\B_{pc}^\Pi,\\
			\B_{xi}&=\left[\proj{w_p}{\B_{ex}}\times\proj{w_c}{\B_{in}}\right]\cap\B_{pc}^\Pi.
		\end{split}
	\end{equation*}	
	In such a case, the largest possible set of controlled behaviour that can be implemented is
	\begin{equation}\label{eq:implementlargest}
		\B_{pc}=\proj{w_p}{\left[\proj{w_p}{\B_d}\times\proj{w_c}{\B_{in}}\right]\cap\B_{pc}^\Pi},
	\end{equation}
	and all controller trajectories that implement $\B_{pc}$ are given by
	\begin{equation}\label{eq:individualcontroller}
		\begin{split}
			\mathfrak{B}_c^j&=\proj{w_c^j}{\mathfrak{B}_{in}}\\
			&=\proj{w_c^j}{\proj{w_c}{\mathfrak{B}_d}\setminus\proj{w_c}{\mathfrak{B}_{ex}}}\\
			&=\proj{w_c^j}{\proj{w_c}{\mathfrak{B}_d}\setminus\proj{w_c}{\mathfrak{B}_{out}}}.
		\end{split}
	\end{equation}
\end{theorem}
\begin{proof}
	See Appendix \ref{appx:proofcontrolsynthesis}.
\end{proof}

We provide a brief explanation of the construction of the various sets in Theorem \ref{theorem:controlsynthesis}. The detailed construction is given in the proof in Appendix \ref{appx:proofcontrolsynthesis}. Since $\B_d$ in \eqref{eq:desirelargest} gives the largest possible set of trajectories that are both within the desired set and admissible through the network, the set $\B_{pc}$, should it exist, must then be a subset of $\proj{w_p}{\B_d}$. All corresponding trajectories of $w_c$ are thus given by $\proj{w_c}{\B_d}$. However, for a given trajectory of $w_c\in\proj{w_c}{\B_d}$, there may be a \emph{set} of trajectories of $w_p$ such that $(w_p,w_c)\in\B_{pc}^{full}$. We call these trajectories the \emph{multiplicities} of a trajectory $w_p$ and we denote the set of all these trajectories as
\begin{equation*}
	\B_p^m=\left\{w_p\in\B_p\mid  \exists w_c\in\proj{w_c}{\B_d},\left(w_p,w_c\right)\in\B_{pc}^{full}\right\}.
\end{equation*}
In general, we have that $\B_p^m\setminus\proj{w_p}{\B_d}\neq\varnothing$, i.e., some trajectories of $w_p\in\proj{w_p}{\B_d}$ would have multiplicities outside of the desired behavioural set. The manifest behaviours of $\B_{in}$ and $\B_{ex}$, i.e., $\proj{w_p}{\B_{in}}$ and $\proj{w_p}{\B_{ex}}$, are all trajectories of $w_p\in\proj{w_p}{\B_d}$ with all multiplicities inside and those with multiplicities outside of $\proj{w_p}{\B_d}$, respectively, $\proj{w_p}{\B_{out}}$ contains all $w_p\not\in\proj{w_p}{\B_d}$ that are admissible through the networks, and $\proj{w_p}{\B_{xi}}$ contains all $w_p\in\proj{w_p}{\B_{ex}}$ with multiplicities in $\proj{w_p}{\B_{in}}$. 

It is immediately obvious that \eqref{eq:exist2} is necessary: it is simply impossible to find a controlled behaviour with $w_f$ being the free variable if it is not free to start with. Furthermore, some manipulations (see the proof in Appendix \ref{appx:proofcontrolsynthesis}) will show that the right hand side of \eqref{eq:exist1} is the projection of $\B_{pc}$ given in \eqref{eq:implementlargest} onto the space $\mathbb{W}_f^\mathbb{T}$. The necessity of \eqref{eq:exist1} then follows because if \eqref{eq:implementlargest} is the largest possible controlled behaviour set, then so it must be for all of its components. The sufficiency of it, however, requires much more delicate proof. 
\begin{remark}\label{remark:dataverify}
	The process of checking \eqref{eq:exist1} can be elusive for model-based representations, but is much easier if the system is represented by data sets. In the ideal case (i.e., where all trajectories of the system are available), \eqref{eq:exist1} reduces to a trajectory selection/elimination problem and the control design is effectively open-loop in the classical sense. However, real data sets are always incomplete and with noise. In such a case, \eqref{eq:exist1} can be checked recursively in a receding-horizon fashion using updated measurements, similar to MPC. The design procedure can also be modified to include a probability measure. Methods to introduce probability descriptions into the framework is currently under investigation.
\end{remark}
\begin{remark}
	While the satisfaction of \eqref{eq:exist2} may seem trivial, its triviality actually comes from the (not necessarily true) modelling assumption that the exogenous inputs admit any arbitrary trajectories. In reality, exogenous inputs may be subject to restraints, hence it is necessary to check whether the chosen free variable $w_f$ is indeed free in $\B_p$. For LTI systems represented by the column span of a Hankel matrix 
	\begin{equation*}\label{eq:Hankelmatrix}
		\mathfrak{H}_L(w)=\begin{bmatrix}
			w(1) & w(2) & \cdots & w(T-L+1)\\
			w(2) & w(3) & \cdots & w(T-L+2)\\
			\vdots & \vdots & \ddots & \vdots\\
			w(L) & w(L+1)&\cdots& w(T)
		\end{bmatrix}.
	\end{equation*}
	constructed by one of its measured trajectories \cite{Willems:2005,Markovsky:2008,Berberich:2020a,Markovsky:2020}, it can be verified by checking whether the submatrix of the Hankel matrix concerning the chosen free variables is of full row rank. For data sets, a probability measure can also be introduced similar to Remark \ref{remark:dataverify}.
\end{remark}

In this theorem we see again the importance of defining the network behaviour in Definition \ref{definition:network} (its importance is more apparent in the proof of Theorem \ref{theorem:controlsynthesis}, see Appendix \ref{appx:proofcontrolsynthesis} for details): it is precisely the network behaviour that builds correspondence between trajectories of $w_p$ and $w_c$ and allows for the construction of various interconnected behaviours from projections. A notable feature in this distributed control design process is that, unlike classical control design where control is based on the inverse of system dynamics, all variables are treated equally and there is no prescribed causality from the to-be-controlled variable to the manipulated variables. They are simply two sets of variables whose trajectories need to be admissible through the system and the controllers are simply restricting the trajectories of the system to a subset that is also a subset of the behaviour describing the desired behaviour rather than inverting the system dynamics in any way.

Another intriguing observation is that in general the largest possible desired set $\proj{w_p}{\B_d}$ \emph{cannot}  be fully implemented. While it is understandable that some trajectories in $\B_p\cap\B_{ps}$ are not implementable because they are not admissible through the network, the interesting result is that even $\proj{w_p}{\B_d}$, whose trajectories all have corresponding trajectories of $w_c$, cannot be fully implemented. This is because, while all trajectories in $\proj{w_c}{\B_d}$ are coming from the desired controlled behaviour, trajectories in $\B_p$ that are admissible with those in $\proj{w_c}{\B_d}$ may not come from $\proj{w_p}{\B_d}$. To make sure that these trajectories are not going to appear, we must eliminate the corresponding trajectories of $w_c$ from $\B_c$, which is why $\B_c^j$ in \eqref{eq:individualcontroller} is also a smaller set than that is depicted in Figure \ref{fig:Desired}, i.e., $\proj{w_c}{\B_d}$. However, by doing so, we also eliminate some trajectories within $\proj{w_p}{\B_d}$ because they are indistinguishable from $w_c$. While $\proj{w_p}{\B_{xi}}$ may ``revive'' some eliminated trajectories in $\proj{w_p}{\B_d}$ (see Appendix \ref{appx:proofcontrolsynthesis}), it is generally not possible to have them all back. Therefore, the largest set of $\B_{pc}$ always satisfies
\begin{equation}\label{eq:implementrange}
	\proj{w_p}{\B_{in}}\subset\B_{pc}\subset\proj{w_p}{\B_d}
\end{equation}
because all trajectories in $\proj{w_p}{\B_{in}}$ are definitely implementable and \eqref{eq:existence} guarantees the non-emptiness of this set (See the proof of Theorem \ref{theorem:controlsynthesis} in Appendix \ref{appx:proofcontrolsynthesis}). Interestingly, the largest and smallest sets in \eqref{eq:implementrange} are obtained when $w_p$ is observable from $w_c$ and when $w_c$ is observable from $w_p$, respectively. This is summarised in the following corollary.

\begin{corollary}\label{corollary:controlsynthesisspecial}
	For the setup in Theorem \ref{theorem:controlsynthesis}, 
	\begin{enumerate}[(i)]
		\item if, additionally, $w_p$ is observable from $w_c$, then the largest implementable behaviour is
		\begin{equation}\label{eq:implementlargestpoc}
			\B_{pc}=\proj{w_p}{\B_d}
		\end{equation} 
		and all controller trajectories are given by 
		\begin{equation}\label{eq:individualcontrollerpoc}
			\B_c^j=\proj{w_c^j}{\B_d};
		\end{equation}
		\item if, additionally, $w_c$ is observable from $w_p$, then the largest implementable behaviour is
		\begin{equation}\label{eq:implementlargestcop}
			\B_{pc}=\proj{w_p}{\B_{in}},
		\end{equation}
		which can be implemented by \eqref{eq:individualcontroller}.
	\end{enumerate}
\end{corollary}
\begin{proof}
	See Appendix \ref{appx:proofcontrolsynthesisspecial}.
\end{proof}

While the case where $w_c$ is observable from $w_p$ is less common, the case where $w_p$ is observable from $w_c$ appears much more frequently (for example, it is generally the case for decentralised control systems and stand-alone systems). Therefore, if one were to search for the controller trajectories, one should begin by checking if either special case in Corollary \ref{corollary:controlsynthesisspecial} applies, in which case the design procedure can be simplified. However, in a distributed control system, $w_p$ and $w_c$ are typically not observable from each other, hence the general conditions in Theorem \ref{theorem:controlsynthesis} should be used.

\section{Conclusion}\label{sec:conclusion}
In this paper, a framework for the analysis and distributed control design of interconnected systems from set-theoretic point of view using behavioural systems theory has been proposed. The network of an interconnected system has been viewed as a dynamical system with its own internal dynamics, which enables the representation of interconnected behaviour to be constructed explicitly from its components, regardless of their respective representations. Furthermore, we have shown that the interconnected behaviour can be completely constructed using the projections of the behaviours of the subsystems from the interconnected system and the behaviour of the network. We have also shown that the same effect can be achieved with any numbers of complete behavioural sets for some subsystems, the projections of the others and the behaviour of the network, allowing for a hybrid platform for model-based/data-driven interconnected systems. Necessary and sufficient conditions for the existence of distributed controllers have been provided and controller behaviours have been constructed explicitly. We believe that this is a more natural view of a dynamical system and is a promising direction for the development of data-driven and hybrid control methods.

\appendix
\renewcommand{\thesubsection}{A.\arabic{subsection}}
\section{Proofs}\label{appx:proofmain}
Before presenting the proofs, we firstly summarise the useful set operations other than the standard operations of $\cap$ and $\cup$ (commutativity, associativity, distributivity and De Morgan's laws) in the following lemma.

\begin{lemma}[Set Operations \cite{Blyth:1975}]\label{lemma:setoperations} \phantom{something}
	
	\begin{enumerate}[(i)]
		\item Let $A$ be a set and let $A_1,A_2\subset A$, then 
		\begin{enumerate}
			\item $A_1\subset A_2$ if and only if $A_1\cap A_2=A_1$ if and only if $A_1\cup A_2=A_2$ if and only if $A_1 \setminus A_2=\varnothing$;
			\item $A_1\cap A_2=\varnothing$ if and only if $A_1\setminus A_2=A_1$;
			\item $A_1=A\setminus A_2$ if and only if $A_1\cap A_2=\varnothing$ and $A_1\cup A_2=A$.
		\end{enumerate}
		\item Let $A^1,A^2,\ldots,A^N$ be $N$ sets with different elements and let $A^i_1,A^i_2\subset A^i, \forall i\in\mathbb{Z}_N^+$, then
		\begin{equation*}
			\bigtimes_{i=1}^{N}\left(A_1^i\cap A_2^i\right)=\left(\bigtimes_{i=1}^{N}A_1^i\right)\cap\left(\bigtimes_{i=1}^{N}A_2^i\right).
		\end{equation*}
	\end{enumerate}
\end{lemma}
Lemma \ref{lemma:setoperations}(ii) is a generalised version of the distributivity of $\times$ over $\cap$, and setting $N=2$ yields a useful identity
\begin{equation*}
	\left(A_1^1\cap A_2^1\right)\times\left(A_1^2\cap A_2^2\right)=\left(A_1^1\times A_1^2\right)\cap\left(A_2^1\times A_2^2\right).
\end{equation*}
Furthermore, we give two auxiliary results that are useful in all subsequent proofs.
\begin{lemma}\label{lemma:projseparate}
	Given $\B^1,\B^2\subset\left(\mathbb{W}_1\times\mathbb{W}_2\right)^\mathbb{T}$ with manifest variable $w=(w_1,w_2)$, we have
	\begin{enumerate}[(i)]
		\item $\proj{w_1}{\B^1\cap\B^2}\subset\proj{w_1}{\B^1}\cap\proj{w_1}{\B^2}$;
		\item $\proj{w_1}{\B^1\cup\B^2}=\proj{w_1}{\B^1}\cup\proj{w_1}{\B^2}$;
		\item $\proj{w_1}{\B^1\setminus\B^2}\supset\proj{w_1}{\B^1}\setminus\proj{w_1}{\B^2}$.
	\end{enumerate}
\end{lemma}
\begin{proof}
	(i) Since $\B^1\cap\B^2\subset\B^1$, we have $\proj{w_1}{\B^1\cap\B^2}\subset\proj{w_1}{\B^1}$. Similar argument can be made to show that $\proj{w_1}{\B^1\cap\B^2}\subset\proj{w_1}{\B^2}$. The two relationships give the result in (i). 
	
	(ii) This is a standard result (see \cite{Imielinski:1984}, for example).
	
	(iii) The set on the left hand side is
	\begin{equation*}
		\begin{split}
			&\proj{w_1}{\B^1\setminus\B^2}=\left\{w^1\mid \exists \ell_2, (w_1,\ell_2)\in\B^1,(w_1,\ell_2)\not\in\B^2\right\}.
		\end{split}
	\end{equation*}
	while the one on the right hand side is
	\begin{equation*}
		\begin{split}
			\proj{w^1}{\B^1}&\setminus\proj{w_1}{\B^2}\\
			=&\left\{w_1\mid \exists \ell_{21}\forall\ell_{22}, (w_1,\ell_{21})\in\B^1,(w_1,\ell_{22})\not\in\B^2\right\}.
		\end{split}
	\end{equation*}
	We see that the latter includes the former with $\ell_{21}=\ell_{22}$.
\end{proof}
\begin{lemma}\label{lemma: projection}
	Given a dynamical system of the form \eqref{eq:twosys}, the following relationships hold:
	\begin{enumerate}[(i)]
		\item $\proj{w^1}{\B}=\B^1\cap\proj{w^1}{\left[\left(\mathbb{W}^1\right)^\mathbb{T}\times\B^2\right]\cap\B^\Pi};$
		\item $\B=\left[\left(\mathbb{W}^1\right)^\mathbb{T}\times\proj{w^2}{\B}\right]\cap\B^\Pi$ if and only if $w^1$ is observable from $w^2$ in $\Sigma$.
	\end{enumerate}
\end{lemma}
\begin{proof}
	(i)  This is straightforward by seeing that the projection of the entire system onto certain spaces is the same as the intersection of the behaviour(s) containing the variables and the rest of the system with the said variables regarded as ``manifest'' variables. 
	
	(ii) Comparing the sets of the right hand side of the expression of $\B$ with the standard construction $\B=\left(\B^1\times\B^2\right)\cap\B^\Pi$, we have that
	\begin{align*}
		\begin{split}
			&\left(\B^1\times\B^2\right)\cap\B^\Pi\\
			=&\left\{(w^1,w^2)\mid w^1\in\B^1, \ w^2\in\B^2, \ (w^1,w^2)\in\B^\Pi\right\},
		\end{split}	\\	
		\begin{split}
			&\left[\left(\mathbb{W}^1\right)^\mathbb{T}\times\proj{w^2}{\B}\right]\cap\B^\Pi \\
			=&\left\{(w^1,w^2)\mid \exists \ell, \ell\in\B^1, \ w^2\in\B^2,\right. \\
			&\qquad\qquad\qquad\quad\left. (\ell,w^2)\in\B^\Pi, \ (w^1,w^2)\in\B^\Pi\right\}.
		\end{split}
	\end{align*}
	The two sets are equal if and only if $\ell=w^1$, and this should hold true for all $w^1$, which is true if and only if $w^1$ is observable from $w^2$.
\end{proof}

\subsection{Proof of Theorem \ref{theorem:behaviourPW}}\label{appx:proofbehaviourPW}
To prove this theorem, we need another auxiliary result, which is stated in the following lemma.
\begin{lemma}\label{lemma:projectioncompare}
	Suppose an interconnected behaviour is of the form \eqref{eq:interconbehaviour}, then
	\begin{enumerate}[(i)]
		\item $\proj{w^i}{\B}\subset\B^i$, $\forall i\in\mathbb{Z}_N^+$;
		\item $\displaystyle \B\subset\bigtimes_{i=1}^{N}\proj{w^i}{\B}$.
	\end{enumerate}
\end{lemma}
\begin{proof}
	(i) This is a direct generalisation from Lemma \ref{lemma: projection}(i).
	
	(ii) Using the definitions of the two sets:
	\begin{equation*}
		\B=\left\{w\mid \forall i,w^i\in\B^i, w\in\B^\Pi\right\}
	\end{equation*}
	and
	\begin{equation*}
		\begin{split}
			&\bigtimes_{i=1}^{N}\proj{w^i}{\B}=\left\{w\mid \forall i \exists \ell_j^i\forall j\neq i, w^i\in\B^i,\ell_j^i\in\B^j,\right. \\ 
			&\left. \qquad\qquad\qquad\quad (\ell_1^i,\ldots,\ell_{i-1}^i,w^i,\ell_{i+1}^i,\ldots,\ell_N^i)\in\B^\Pi\right\},
		\end{split}
	\end{equation*}
	we see that the former is the latter with an extra condition: $\ell_j^i=w^j, \forall i, j$. 
\end{proof}

Now we are ready to prove the theorem. 

(i) From Lemma \ref{lemma: projection}(i), we have
\begin{equation*}
	\begin{split}
		&\proj{w^i}{\B}\\
		=&\B^i\cap\proj{w^i}{\left[\left(\bigtimes_{j=1}^{i-1}\B^j\right)\times\left(\mathbb{W}^i\right)^\mathbb{T}\times\left(\bigtimes_{j=i+1}^{N}\B^j\right)\right]\cap\B^\Pi}\\
		\coloneqq &\B^i\cap\B_\pi^i
	\end{split}
\end{equation*} 
It follows from Lemma \ref{lemma:setoperations}(iii) that
\begin{equation*}
	\begin{split}
		&\left(\bigtimes_{i=1}^{N}\proj{w^i}{\B}\right)\cap\B^\Pi=\left(\bigtimes_{i=1}^{N}\B^i\cap\B_\pi^i\right)\cap\B^\Pi\\
		=&\left(\bigtimes_{i=1}^{N}\B^i\right)\cap\left(\bigtimes_{i=1}^{N}\B_\pi^i\right)\cap\B^\Pi=\B\cap\left(\bigtimes_{i=1}^{N}\B_\pi^i\right).
	\end{split}
\end{equation*}
According to Lemma \ref{lemma:setoperations}(i), it suffices to prove that $\B\subset\bigtimes_{i=1}^{N}\B_\pi^i$. Obviously, for all $i$, we have
\begin{align*}
	\B_\pi^i=&\proj{w^i}{\left[\left(\bigtimes_{j=1}^{i-1}\B^j\right)\times\left(\mathbb{W}^i\right)^\mathbb{T}\times\left(\bigtimes_{j=i+1}^{N}\B^j\right)\right]\cap\B^\Pi}\\
	\supset&\proj{w^i}{\left[\left(\bigtimes_{j=1}^{i-1}\B^j\right)\times\B^i\times\left(\bigtimes_{j=i+1}^{N}\B^j\right)\right]\cap\B^\Pi}\\
	=&\proj{w^i}{\left(\bigtimes_{i=1}^{N}\B^i\right)\cap\B^\Pi}=\proj{w^i}{\B}.
\end{align*}
It then follows from Lemma \ref{lemma:projectioncompare}(ii) that
\begin{equation*}
	\begin{split}
		\bigtimes_{i=1}^{N}\B_\pi^i\supset\bigtimes_{i=1}^{N}\proj{w^i}{\B}\supset\B.
	\end{split}
\end{equation*}
(ii) Note that according to Lemma \ref{lemma:projectioncompare}(ii), $\proj{w^i}{\B}\subset\B^i,\ \forall i$. Combining with the result in (i), it then follows that
\begin{align*}
	\B&=\left(\bigtimes_{i=1}^{N}\proj{w^i}{\B}\right)\cap\B^\Pi\\
	&\subset\left[\left(\bigtimes_{i=1}^{n}\B^i\right)\times\left(\bigtimes_{i=n+1}^{N}\proj{w^i}{\B}\right)\right]\cap\B^\Pi\\
	&\subset\left(\bigtimes_{i=1}^{N}\B^i\right)\cap\B^\Pi=\B.
\end{align*}
This completes the proof of Theorem \ref{theorem:behaviourPW}.
\subsection{Proof of Theorem \ref{theorem:controlsynthesis}}\label{appx:proofcontrolsynthesis}
(\emph{only if}): Suppose conditions in \eqref{eq:objectives} hold, then \eqref{eq:exist2} is immediate because 
\begin{equation*}
	\proj{w_f}{\B_p}\supset\proj{w_f}{\B_p\cap\B_{ps}}\supset\proj{w_f}{\B_{pc}}=\mathbb{W}_f^\mathbb{T}.
\end{equation*}
To show \eqref{eq:exist1}, we begin by explaining the construction of the various sets in the theorem. In the situation where the multiplicity set $\B_p^m\setminus\proj{w_p}{\B_d}\neq\varnothing$, the corresponding trajectories in $w_c$ need to be excluded from control design because all multiplicities are indistinguishable from $w_c$. The remaining valid controller behaviour can be constructed in the following way:
\begin{enumerate}
	\item Find all trajectories of $w_c$ projected from integrating a dynamical system containing all trajectories of $w_p$ belonging to $\B_p$ but not to $\proj{w_p}{\B_d}$ into the system. This gives $\proj{w_c}{\B_{out}}$;
	\item All excluded trajectories of $w_p$ can be found by projecting all of $w_c$ found from the previous step to $w_p$ through the system and intersecting with $\proj{w_p}{\B_d}$. This gives the set $\proj{w_p}{\B_{ex}}$;
	\item The largest possible set of the controller behaviour $\B_c$ is hence the subset of $\B_c^\Pi\cap\B_{cr}$ containing all trajectories of $w_c$ projected from integrating $\proj{w_p}{\B_d}$, excluding $\proj{w_p}{\B_{ex}}$, into the network, which is precisely $\proj{w_c}{\B_{in}}$.
\end{enumerate}
While the above procedure gives $\proj{w_c}{\B_{in}}$, integrating this into the network may result in a corresponding behaviour of $w_p$ that is larger than $\proj{w_p}{\B_{in}}$ and a portion of it may end up in $\proj{w_p}{\B_{ex}}$. This set of trajectories, should they exist, are also implementable, and they are given by $\proj{w_p}{\B_{xi}}$. Therefore, if there exists $\B_c$ such that \eqref{eq:objectives} holds, we must have $\B_c\subset\proj{w_c}{\B_{in}}$, hence 
\begin{equation*}
	\B_{ps}\subset\proj{w_p}{\B_{in}}\cup\proj{w_p}{\B_{xi}}.
\end{equation*}
Therefore, according to Lemma \ref{lemma:projseparate}(ii), we have
\begin{equation}\label{eq:regioncombinefree}
	\begin{split}
		&\proj{w_f}{\B_{in}}\cup\proj{w_f}{\B_{xi}}\\
		=&\proj{w_f}{\proj{w_p}{\B_{in}}}\cup\proj{w_f}{\proj{w_p}{\B_{xi}}}\\
		=&\proj{w_f}{\proj{w_p}{\B_{in}}\cup\proj{w_p}{\B_{xi}}}\supset\proj{w_f}{\B_{ps}}.
	\end{split}
\end{equation}
Then \eqref{eq:exist1} follows due to \eqref{eq:DOF}. Furthermore, according to Corollary \ref{corollary:twosystemproj}, we have
\begin{equation*}
	\B_{in}=\left[\left(\proj{w_p}{\B_d}\setminus\proj{w_p}{\B_{ex}}\right)\times\proj{w_c}{\B_{in}}\right]\cup\B_{pc}^\Pi.
\end{equation*}
As a result,
\begin{equation}\label{eq:regioncombinemani}
	\begin{split}
		&\proj{w_p}{\B_{in}}\cup\proj{w_p}{\B_{xi}}\\
		=&\proj{w_p}{\B_{in}\cup\B_{xi}}\\
		=&\proj{w_p}{\left[\proj{w_p}{\B_d}\times\proj{w_c}{\B_{in}}\right]\cap\B_{pc}^\Pi},
	\end{split}
\end{equation}
which gives the largest possible implementable set \eqref{eq:implementlargest}. This completes the \emph{only if} part of the proof.

(\emph{if}): Suppose that conditions in \eqref{eq:existence} hold. We begin by showing that $\B_{in}\neq\varnothing$. In other words, the desired controller behaviour is guaranteed to exist. 

Suppose $\B_{in}=\varnothing$, then by construction, $\proj{w_f}{\B_{in}}=\proj{w_f}{\B_{xi}}=\varnothing$. According to \eqref{eq:exist1}, we must have that
\begin{equation*}
	\proj{w_f}{\B_p\setminus\proj{w_p}{\B_{in}}\setminus\proj{w_p}{\B_{xi}}}=\varnothing.
\end{equation*}
By assumption, $\proj{w_p}{\B_{in}}=\proj{w_p}{\B_{xi}}=\varnothing$, meaning that $\proj{w_f}{\B_p}=\varnothing$. This contradicts with \eqref{eq:exist2}, hence $\B_{in}\neq\varnothing$.

Now, note that $\proj{w_c}{\B_{out}}\subset\B_c^\Pi\cap\B_{cr}$, then according to Corollary \ref{corollary:twosystemproj} and Lemma \ref{lemma: projection}(i), we have
\begin{align}
	&\proj{w_c}{\B_{in}}\cap\proj{w_c}{\B_{out}} \nonumber\\
	=&\proj{w_c}{\left[\left(\proj{w_p}{\B_d}\setminus\proj{w_p}{\B_{ex}}\right)\times\proj{w_c}{\B_{out}}\right]\cap\B_{pc}^\Pi} \nonumber\\
	=&\proj{w_c}{\B_{ex}\setminus\B_{ex}}=\varnothing. \label{eq:impoutexclude}
\end{align}
This means that
\begin{equation*}
	\begin{split}
		&\left[\left(\B_p\setminus\proj{w_p}{\B_d}\right)\times\proj{w_c}{\B_{in}}\right]\cap\B_{pc}^\Pi\\
		=&\left[\left(\B_p\setminus\proj{w_p}{\B_d}\right)\times\left(\proj{w_c}{\B_{in}}\cap\B_c^\Pi\cap\B_{cr}\right)\right]\cap\B_{pc}^\Pi\\
		=&\B_{out}\cap\left[\left(\B_p\setminus\proj{w_p}{\B_d}\right)\times\proj{w_c}{\B_{in}}\right]\cap\B_{pc}^\Pi\\
		=&\left[\left(\B_p\setminus\proj{w_p}{\B_d}\right)\times\left(\proj{w_c}{\B_{in}}\cap\proj{w_c}{\B_{out}}\right)\right]\cap\B_{pc}^\Pi\\
		=&\varnothing.
	\end{split}
\end{equation*}
Therefore,
\begin{align}\label{eq:forwardinclusion}
	&\proj{w_p}{\left[\B_p\times\proj{w_c}{\B_{in}}\right]\cap\B_{pc}^\Pi} \nonumber\\
	=&\proj{w_p}{\left[\left(\left(\B_p\setminus\proj{w_p}{\B_d}\right)\cup\proj{w_p}{\B_d}\right)\times\proj{w_c}{\B_{in}}\right]\cap\B_{pc}^\Pi} \nonumber\\
	=&\proj{w_p}{\left[\proj{w_p}{\B_d}\times\proj{w_c}{\B_{in}}\right]\cap\B_{pc}^\Pi}.
\end{align}
This means that $\B_{pc}$ given in \eqref{eq:implementlargest} can be implemented by choosing
\begin{equation}\label{eq:distributedcontroller}
	\B_c=\proj{w_c}{\B_{in}}.
\end{equation}
It is easy to see that $\B_{pc}$ satisfies \eqref{eq:implementability} because
\begin{align*}
	\proj{w_p}{\left[\proj{w_p}{\B_d}\times\proj{w_c}{\B_{in}}\right]\cap\B_{pc}^\Pi}\subset\proj{w_p}{\B_d}\subset\B_p\cap\B_{ps}.
\end{align*}
Furthermore, similar to the derivation in \eqref{eq:regioncombinefree}, \eqref{eq:exist1} means that
\begin{align*}
	&\proj{w_f}{\B_{in}}\cup\proj{w_f}{\B_{xi}}\\
	=&\proj{w_f}{\B_{in}}\cup\proj{w_f}{\B_{xi}}\cup\proj{w_f}{\B_p\setminus\proj{w_p}{\B_{in}}\setminus\proj{w_p}{\B_{xi}}}\\
	=&\proj{w_f}{\B_p}
\end{align*}
Since $\proj{w_f}{\B_p}=\mathbb{W}_f^\mathbb{T}$ due to \eqref{eq:exist2} and $\proj{w_f}{\B_{in}}\cup\proj{w_f}{\B_{xi}}$ is equivalent to $\B_{pc}$ in \eqref{eq:implementlargest} according to \eqref{eq:regioncombinemani}, we achieve \eqref{eq:DOF}.

We now show that choosing individual controllers as \eqref{eq:individualcontroller} will result in \eqref{eq:distributedcontroller}. Since $\B_c$ is of the form \eqref{eq:interconbehaviour} and $\proj{w_c}{\B_{in}}\subset\B_c^\Pi$, we have
\begin{equation*}
	\B_c=\left(\bigtimes_{j=1}^{N_c}\B_c^j\right)\cap\B_c^\Pi=\left(\bigtimes_{j=1}^{N_c}\proj{w_c^j}{\B_c}\right)\cap\B_c^\Pi.
\end{equation*}
By choosing \eqref{eq:individualcontroller}, we can construct $\B_c$ according to Theorem \ref{theorem:behaviourPW}(i) as
\begin{align*}
	\B_c&=\left(\bigtimes_{j=1}^{N_c}\proj{w_c^j}{\B_{in}}\right)\cap\B_c^\Pi\\
	&=\left(\bigtimes_{j=1}^{N_c}\proj{w_c^j}{\proj{w_c}{\B_{in}}}\right)\cap\B_c^\Pi=\proj{w_c}{\B_{in}}.
\end{align*}
Finally, we show the equivalence of the three representations in \eqref{eq:individualcontroller}. We firstly show that
\begin{equation}\label{eq:controlminus}
	\proj{w_c}{\B_{in}}=\proj{w_c}{\B_d}\setminus\proj{w_c}{\B_{ex}},
\end{equation}
which, according to Lemma \ref{lemma:setoperations}(i), is equivalent to proving 
\begin{subequations}
	\begin{align}
		\proj{w_c}{\B_{in}}\cap\proj{w_c}{\B_{ex}}&=\varnothing, \label{eq:controlminus1}\\
		\proj{w_c}{\B_{in}}\cup\proj{w_c}{\B_{ex}}&=\proj{w_c}{\B_d}. \label{eq:controlminus2}
	\end{align}
\end{subequations}
Since $\proj{w_c}{\B_{ex}}\subset\proj{w_c}{\B_{out}}$, \eqref{eq:controlminus1} can be directly deduced from \eqref{eq:impoutexclude}. We prove \eqref{eq:controlminus2} by contradiction. To begin with, it is easy to see that
\begin{equation*}
	\proj{w_c}{\B_{in}}\cup\proj{w_c}{\B_{ex}}\subset\proj{w_c}{\B_d}\cup\proj{w_c}{\B_d}=\proj{w_c}{\B_d}.
\end{equation*}
Suppose that there exists a subset of $\proj{w_c}{\B_d}$, call it the residual set $\proj{w_c}{\B_{res}}$, such that
\begin{equation*}
	\proj{w_c}{\B_{res}}=\proj{w_c}{\B_d}\setminus\proj{w_c}{\B_{in}}\setminus\proj{w_c}{\B_{ex}}
\end{equation*}
and suppose that $\proj{w_c}{\B_{res}}\neq\varnothing$. Since $\proj{w_p}{\B_{ex}}\subset\proj{w_p}{\B_d}$ and
\begin{align*}
	&\proj{w_p}{\B_{in}}\\
	=&\proj{w_p}{\B_d}\setminus\proj{w_p}{\B_{ex}}\\
	&\qquad\qquad\cap\proj{w_p}{\left[\proj{w_p}{\B_d}\times\left(\B_c^\Pi\cap\B_{cr}\right)\right]\cap\B_{pc}^\Pi}\\
	=&\proj{w_p}{\B_d}\setminus\proj{w_p}{\B_{ex}}\cap\proj{w_p}{\B_d}\\
	=&\proj{w_p}{\B_d}\setminus\proj{w_p}{\B_{ex}},
\end{align*}
it follows that $\proj{w_p}{\B_{in}}\cup\proj{w_p}{\B_{ex}}=\proj{w_p}{\B_d}$ and that $\proj{w_p}{\B_{in}}\cap\proj{w_p}{\B_{ex}}=\varnothing$. Therefore, for all $w_c\in\proj{w_c}{\B_d}$ there must exist at least one $w_p$ that belongs to either $\proj{w_p}{\B_{in}}$ or $\proj{w_p}{\B_{ex}}$ such that $(w_p,w_c)\in\B_{pc}^\Pi$. We first show that for $w_c\in\proj{w_c}{\B_{res}}$, the corresponding trajectories of $w_p$ cannot come from $\proj{w_p}{\B_{in}}$, i.e.,
\begin{equation*}
	\left[\proj{w_p}{\B_{in}}\times\proj{w_c}{\B_{res}}\right]\cap\B_{pc}^\Pi=\varnothing.
\end{equation*}
Notice that
\begin{align*}
	&\B_d\setminus\B_{in}\\
	=&\B_d\setminus\B_{in}\cap\B_d\\
	=&\left[\proj{w_p}{\B_{ex}}\times\left(\B_c^\Pi\cap\B_{cr}\right)\right]\cap\B_{pc}^\Pi\\
	&\qquad\qquad\qquad\qquad\quad\cap\left[\proj{w_p}{\B_d}\times\proj{w_c}{\B_d}\right]\cap\B_{pc}^\Pi\\
	=&\left[\proj{w_p}{\B_{ex}}\times\proj{w_c}{\B_d}\right]\cap\B_{pc}^\Pi.
\end{align*}
As a result,
\begin{equation*}
	\begin{split}
		&\left[\proj{w_p}{\B_{in}}\times\proj{w_c}{\B_{res}}\right]\cap\B_{pc}^\Pi\\
		\subset&\left[\proj{w_p}{\B_{in}}\times\left(\proj{w_c}{\B_d}\setminus\proj{w_c}{\B_{in}}\right)\right]\cap\B_{pc}^\Pi\\
		=&\left(\left[\proj{w_p}{\B_{in}}\times\proj{w_c}{\B_d}\right]\cap\B_{pc}^\Pi\right)\setminus\B_{in}\\
		=&\left(\left[\left(\proj{w_p}{\B_d}\setminus\proj{w_p}{\B_{ex}}\right)\times\proj{w_c}{\B_d}\right]\cap\B_{pc}^\Pi\right)\setminus\B_{in}\\
		=&\B_d\setminus\B_{in}\setminus\left(\left[\proj{w_p}{\B_{ex}}\times\proj{w_c}{\B_d}\right]\cap\B_{pc}^\Pi\right)=\varnothing.
	\end{split}
\end{equation*}
Therefore, all corresponding trajectories of $w_p\in\proj{w_p}{\B_d}$ for $w_c\in\proj{w_c}{\B_{res}}$ must belong to $\proj{w_p}{\B_{ex}}$. On the other hand, all trajectories in $\proj{w_p}{\B_{ex}}$ have multiplicities in $\B_p\setminus\B_{ps}$, it follows that
\begin{equation*}
	\begin{split}
		\proj{w_c}{\B_{res}}&\subset\proj{w_c}{\left[\left(\B_p\setminus\proj{w_p}{\B_d}\right)\times\left(\B_c^\Pi\cap\B_{cr}\right)\right]\cap\B_{pc}^\Pi}\\
		&=\proj{w_c}{\B_{out}},
	\end{split}
\end{equation*}
but $\proj{w_c}{\B_{res}}\subset\proj{w_c}{\B_d}$ by definition. Therefore,
\begin{equation*}
	\begin{split}
		\proj{w_c}{\B_{res}}&\subset\proj{w_c}{\B_d}\cap\proj{w_c}{\B_{out}}\\
		&=\proj{w_c}{\left[\proj{w_p}{\B_d}\times\proj{w_c}{\B_{out}}\right]\cap\B_{pc}^\Pi}\\
		&=\proj{w_c}{\B_{ex}}.
	\end{split}
\end{equation*}
But $\proj{w_c}{\B_{res}}\not\subset\proj{w_c}{\B_{ex}}$ by construction. This contradiction means  that $\proj{w_c}{\B_{res}}=\varnothing$. Therefore, \eqref{eq:controlminus2}, hence \eqref{eq:controlminus}, is satisfied. This establishes the equivalence between the first two representations in \eqref{eq:individualcontroller} because $\proj{w_c^j}{\B_{in}}=\proj{w_c^j}{\proj{w_c}{\B_{in}}}$. Furthermore,
\begin{align}\label{eq:controlminusalter}
	&\proj{w_c}{\B_d}\setminus\proj{w_c}{\B_{ex}} \nonumber\\
	=&\proj{w_c}{\B_d}\setminus\proj{w_c}{\left[\left(\B_p\cap\B_{ps}\right)\times\proj{w_c}{\B_{out}}\right]\cap\B_{pc}^\Pi}\nonumber\\
	=&\proj{w_c}{\B_d}\setminus\left[\proj{w_c}{\left[\left(\B_p\cap\B_{ps}\right)\times\mathbb{W}_c^\mathbb{T}\right]\cap\B_{pc}^\Pi}\cap\proj{w_c}{\B_{out}}\right]\nonumber\\
	=&\left(\proj{w_c}{\B_d}\setminus\left[\proj{w_c}{\left[\left(\B_p\cap\B_{ps}\right)\times\mathbb{W}_c^\mathbb{T}\right]\cap\B_{pc}^\Pi}\right]\right)\nonumber\\
	&\qquad\qquad\qquad\qquad\qquad\quad\cup\left[\proj{w_c}{\B_d}\setminus\proj{w_c}{\B_{out}}\right]\nonumber\\
	=&\proj{w_c}{\B_d}\setminus\proj{w_c}{\B_{out}}.
\end{align}
\eqref{eq:controlminus} and \eqref{eq:controlminusalter} give the equivalences in \eqref{eq:individualcontroller}.

Notice that \eqref{eq:individualcontroller} is the smallest controller behavioural set to achieve \eqref{eq:distributedcontroller}. $\B_c^j$ may contain other trajectories but they must not be admissible through $\B_c^\Pi$. This completes the \emph{if} part, thus the proof of Theorem \ref{theorem:controlsynthesis}.

\subsection{Proof of Corollary \ref{corollary:controlsynthesisspecial}}\label{appx:proofcontrolsynthesisspecial}
(i) For this case, we show that $\proj{w_p}{\B_{ex}}=\varnothing$, hence deducing that $\proj{w_p}{\B_{in}}=\proj{w_p}{\B_d}$. \eqref{eq:implementlargest} and \eqref{eq:individualcontroller} then specialise to \eqref{eq:implementlargestpoc} and \eqref{eq:individualcontrollerpoc}, respectively. According to Lemma \ref{lemma: projection}(ii), since $w_p$ is observable from $w_c$, we have
\begin{equation*}
	\begin{split}
		\proj{w_p}{\B_{ex}}&=\left(\B_p\cap\B_{ps}\right)\cap\proj{w_p}{\left[\mathbb{W}_p^\mathbb{T}\times\proj{w_c}{\B_{out}}\right]\cap\B_{pc}^\Pi}\\
		&=\B_p\cap\B_{ps}\cap\proj{w_p}{\B_{out}}.
	\end{split}
\end{equation*}
Now, notice that
\begin{equation*}
	\begin{split}
		\proj{w_p}{\B_{out}}&=\proj{w_p}{\left[\left(\B_p\setminus\B_{ps}\right)\times\left(\B_c^\Pi\cap\B_{cr}\right)\right]\cap\B_{pc}^\Pi}\\
		&\subset\B_p\setminus\B_{ps}
	\end{split}
\end{equation*}
it follows that
\begin{equation*}\label{eq:emptycaseI}
	\proj{w_p}{\B_{ex}}\subset\left(\B_p\cap\B_{ps}\right)\cap\left(\B_p\setminus\B_{ps}\right)=\varnothing.
\end{equation*}

(ii) Similar to the previous case, we show that $\B_{xi}=\varnothing$ for this case. \eqref{eq:implementlargestcop} then follows from \eqref{eq:regioncombinemani}. According to Lemma \ref{lemma: projection}(ii), since $w_c$ is observable from $w_p$, we have
\begin{align*}
	\B_{xi}&=\left[\proj{w_p}{\B_{ex}}\times\mathbb{W}_c^\mathbb{T}\right]\cap\left[\mathbb{W}_p^\mathbb{T}\times\proj{w_c}{\B_{in}}\right]\cap\B_{pc}^\Pi\\
	&=\B_{ex}\cap\left[\mathbb{W}_p^\mathbb{T}\times\proj{w_p}{\B_{in}}\right]\cap\B_{pc}^\Pi\\
	&=\left[\proj{w_p}{\B_d}\times\left(\proj{w_c}{\B_{out}}\cap\proj{w_c}{\B_{in}}\right)\right]\cap\B_{pc}^\Pi.
\end{align*}
The emptiness then follows from \eqref{eq:impoutexclude}.

\bibliographystyle{elsarticle-num}
\bibliography{YanBaoHuang}

\end{document}